\documentstyle[epsf]{l-aa}
\newcommand{\ms}{$M_{\odot}$}

\newcommand{\kms}{km\,s$^{-1}$}

\newcommand{\cmc}{cm$^{-3}$}
\newcommand{\ea}{et~al.~}
\newcommand{\hi}{H\,{\sc i}}
\newcommand{\hii}{H\,{\sc ii}}

\begin{document}

\title{{\it Letter to the Editor}\\
Bow-shock induced star formation in the LMC?}

\author{K.S. de Boer\inst{1}
\and J.M. Braun\inst{1}
\and A. Vallenari\inst{2}
\and U. Mebold\inst{3}
}

\institute{
Sternwarte, Univ. Bonn, Auf dem H\"ugel 71, D-53121 Bonn, Germany
\and
Osservatorio Astronomico di Padova, Vicolo Osservatorio 5, I-35122 Padova
\and
Radioastronomisches Institut, Univ. Bonn, Auf dem H\"ugel 71, 
	D-53121 Bonn, Germany
}

\date{ received 11 Sept. 1997; accepted 30 Oct. 1997 }

\thesaurus{08.06.2; 11.05.2; 11.11.1; 11.13.1}

\offprints{deboer@astro.uni-bonn.de}

\maketitle

\markboth{K.S. de Boer et al.,  LMC bow-shock star formation}
	{K.S. de Boer et al., LMC bow-shock star formation}

\begin{abstract}

The structure of supergiant shells, in particular of LMC~4, 
is hard to explain with stochastic self-propagating star formation.  
A series of supergiant structures lies along the outer edge of the LMC 
and form a sequence increasing clockwise in age. 
We have considered the rotation of the LMC and its motion 
through the halo of the Milky Way and propose that 
these structures find their origin in star formation 
induced in the bow-shock formed at the leading edge of the LMC. 
Due to the rotation of the LMC these structures then move aside. 

\end{abstract}

\section{Introduction}

The star formation history of dwarf galaxies is thought to be erratic. 
Since the LMC is nearby and visible at only modest inclination, 
we have a good overview of its structure. 
The LMC contains numerous \hii\ regions and 
filamentary H$\alpha$-light emitting bubbles. 
A few regions appear to have coherent sets of \hii\ regions, leading 
Goudis \& Meaburn (1978) to propose the existence of supergiant shells. 
The creation scenario for supergiant shells was right from the start one of 
stochastic self-propagating star formation, SSPSF  
(Feitzinger \ea 1981). 

The most prominent structures in the LMC have received most attention, 
of course. 
The young stars are bright and easy to measure so that for many 
young clusters and associations ages have been determined. 
Since older star groups in the same field as the young ones 
overlay each other in the Hertzsprung-Russell diagram (HRD), 
their age is more difficult to determine. 
The age of older groups has largely been derived from 
inconspicuous fields (see reviews by Vallenari 1996, Olszewski \ea 1996).

If the concept of SSPSF were valid, in particular the larger structures 
should show a gradient in age of the stars, 
being very young at the edges and being older toward the inside. 
Supergiant shell LMC~4 with a diameter of about 1 kpc 
seemed to be a prime case for such an investigation. 
As it turned out, 
the age structure in the interior predicted from SSPSF (Dopita \ea 1985) 
is in conflict with the age of the stars at the edge as well as 
that of stars in the interior 
(Vallenari \ea 1993; Will \ea 1996; Braun \ea 1997). 
The very fact that the stars in the entire interior of LMC~4 
are essentially of the same age (Braun \ea 1997) 
leads us to look for a large scale trigger for the formation 
of stars in these large structures. 

\section{Large scale structure and motion of the LMC}

The most prominent young structures of the LMC are: 
the 30 Doradus region with young stars and brilliant \hii\ emission 
(see e.g. Walborn 1984); 
supergiant shell LMC~4 with a diameter of $\sim$1 kpc to the N 
of 30 Dor (Meaburn 1980); 
the region south of 30 Dor being dark in the visual and X-ray 
(Blondiau \ea 1997) but very bright in the infrared (see Schwering 1988);
the relatively sharp boundary of the \hi\ gas toward the SE of the 
LMC (see Mathewson \& Ford 1984). 
Do these features have a common explanation?

The LMC as a (dwarf)galaxy moves in an orbit around the Milky Way. 
Various models for its motion exist, and the common opinion is that the LMC 
is (as part of the Magellanic System) at present close to perigalacticon. 
The previous closest approach to the disk was about 1.5 Gyr ago 
(see e.g.\, Heller \& Rohlfs 1994).
The LMC and the SMC move about each other and had a closest approach 
$\sim$200 Myr ago.
The motion of the LMC is directed toward the East, 
i.e., towards the galactic plane. 

In its motion through the halo of the Milky Way, 
the LMC gas obviously is getting compressed at the leading edge. 
This then explains the sharp \hi\ boundary on the SE side, 
while at the trailing side the gas gets diffuse and 
forms the Magellanic Stream (Mathewson \& Ford 1984).  

\section{An external star formation trigger}

We propose the following scenario. 
Star formation is triggered in the gas being compressed at the 
leading edge due to the bow-shock of the LMC. 
The favoured location is at the SE side, where the cumulative effect 
of the LMC space velocity and the velocity of LMC rotation is largest. 
Since the bow-shock compresses large areas, this will lead to 
star formation on a large scale and thus to large structures. 
Because of the clock-wise rotation, the material at the leading edge will, 
in time, move away to the side. 
Thus, when looking at superstructures away from the leading edge, 
we expect to find a progression in the age of such structures in the 
direction of the rotation. 
Substantial evidence is now available to support this scenario.

\section{Velocity, rotation, and inclination of the LMC}

The radial velocity of the LMC is well determined at +274 \kms. 
This value is based both on stellar radial velocities as well as 
on velocities of \hi. 

The rotation curve of the LMC in the {\it radial\,} sense can be derived 
from the positional variations in \hi\ radial velocities. 
Two thorough investigations (Meatheringham \ea 1988; Luks \& Rohlfs 1992) 
document that the radial rotation curve has an amplitude of $\sim 60$ \kms.  

The lateral motion of the LMC was recently derived by 
Kroupa \& Bastian (1997) from an analysis of proper motion 
measurements of stars in the field of the LMC by Hipparcos. 
They derive a proper motion of 0.00195\arcsec /yr.
The direction and value of the motion is in line with that 
from the models for the spatial motion of the Magellanic System as a whole 
(see Heller \& Rohlfs 1994). 

The rotation velocity as seen {\it in projection\,} was investigated 
by Kroupa \& Bastian (1997) 
based on the positional variation of the proper motions. 
They find a clockwise rotation of $58 \pm 58$ \kms\ 
refered to a radius of 1.3 kpc. 
The result is close to the limit of the accuracy of the data 
and no information about the behaviour of the rotation curve with radius 
is available.  
Since the LMC is from our vantage point inclined by about 30$^{\rm o}$ 
(see data compiled by Westerlund, 1997, Table 3.5), 
the Kroupa \& Bastian value is an average from those parts of the LMC 
showing the full rotation as a tangential one and those where 
the full rotation is divided over tangential and radial rotation. 
However, the uncertainty of the astrometric data do not allow 
to pursue this in detail. 
Jones \ea (1994) found from an outlying field a rotation of 180 \kms. 

We must note that the astrometry for the determination of the rotation 
used only the brightest stars, 
which are among the most massive and thus the youngest.  
The radial component of rotation is derived from the  radial velocities 
of \hi\ gas.  
Both determinations therefore refer to the motion of 
the `young' component of the LMC. 
Whether and how the old star complexes of the LMC, such as the bar, 
participate in the rotation is at present unknown. 

New Australia Telescope \hi\ 21 cm data (Kim \ea  1997) 
show the LMC as an essentially circular gas disk, 
suggesting a smaller inclination. 
With a small inclination of 22$^{\rm o}$ (Kim \ea 1997) 
the full rotation velocity must 
be well over 100 \kms\ in order to explain the radial rotation curve. 

Whatever may be the case, based on all discussed indications we will assume 
for this paper a small inclination and a full rotation velocity of 
$\sim150$ \kms\ 
for all positions at distances further out than 1.5 kpc from the centre.

\section{Total speed and 
gas compression at leading edge}

The LMC moves through the halo of the Milky Way with a 
galactocentric velocity of 265 \kms, 
mostly directed tangentially to our line of sight to the LMC 
(Kroupa \& Bastian 1997). 
In addition to this, the LMC rotates with a speed of $\sim$150 \kms\ 
(see above), 
a rotation such that the South side of the LMC moves in the same direction 
as the motion of the LMC. 
LMC gas ahead of the southern side will be exposed to the sum of 
these velocities, being 415 \kms, in the Milky Way frame. 
Furthermore, the galactic halo itself rotates against the motion 
of the LMC, but at this distance with most likely a small speed. 
We will assume $\sim$50 \kms. 
Thus the total velocity difference between the halo gas and the LMC gas 
is $\sim$465 \kms. 

Due to the velocity difference of $\sim$465 \kms\ 
the LMC gas will get compressed. 
What features are expected?

At such velocities, the interaction with the tenuous halo gas
must lead to friction and thus heating. 
We expect to see a shock indeed, in which the density and temperature 
will be high enough to produce X-rays. 

The gas density of the halo of the Milky Way is not well defined. 
The C\,{\sc iv} absorption line profiles seen in Magellanic Cloud star 
spectra suggest a gas density of $10^{-4}$ \cmc\ at $z = 10$ kpc 
at perhaps 10$^5$ K (Savage \& de Boer 1981), or a higher density 
if the temperature were higher. 

A similar density is found by Weiner \& Williams (1996) who observed 
H$\alpha$ emission at the leading edge of Magellanic Stream cloud MS IV. 
They derive a density of about 10$^{-4}$ \cmc\ for the halo, 
based on the assumption that this emission is fed by the energy influx 
into the frontal area of MS IV due to interaction with the halo plasma. 
Adopting a temperature of $10^6$ to $10^7$ K for the halo, 
they find the pressure in the halo at the distance of the Magellanic Clouds 
to be of order $nT = 10^2$ to $10^3$ K \cmc. 
These are rather moderate pressures and would not lead to enhanced star 
formation activity.

\begin{table}
\caption[]{Parameters for age and position of superstructures 
along the edge of the LMC}
\begin{tabular}{lrrl}
\hline
Name			& Distance$^a$	&  Age  & Ref.\\
			& (kpc)		& (Myr) & age\\
\hline
Dark cloud		&  0		& $< 0$ \\
N 159 			&  0.5		& $< 3$	& 1\\
30 Dor			&  1.1		&   3-5 & 2\\
LMC 4 (Sh III)		&  3.0 		&  9-16 & 3\\
NGC 1818 and field	&  6.0		& 20-40 & 4\\
Field near NGC 1783	& 6.7		& 20-50 & 5\\
\hline
\end{tabular}

\noindent
$^a$ All distances are related to the brightest IRAS far-IR emission in 
the dark cloud south of 30\,Dor\\
Refs.: 1 = from the CMD in Deharveng \& Caplan (1991);
2 = De Marchi \ea (1993); see also the data in Parker \& Garmany (1993);
3 = Braun \ea (1997); 
4 = Will \ea (1995) and Hunter \ea (1997);
5 = Cole \ea (1997)
\end{table}


The pressure at the south-eastern edge of the LMC has been derived by 
Blondiau \ea (1997) from X-ray emission observed with the ROSAT telescope. 
They find $n_{\rm e} T_{\rm e} = 1.5\ 10^5$ K \cmc\ from the X-ray spectrum.
This pressure is very large even compared to estimates for the pressure 
in our Galaxy, which are between $10^4$ K \cmc\ and 3 $10^4$ K \cmc. 
Since the spatial average of the pressure is coupled to the gravitational 
potential of a galaxy, the typical pressure in the LMC 
will certainly be less than that in our Galaxy. 
Therefore we can safely state that the pressure in the X-ray emitting region 
south-east of 30\,Dor is more than 10 times as high as on average 
in the rest of the LMC. 
This high pressure region is located near the leading edge of the LMC, 
with a relative velocity with respect to the halo gas 
of about 465 \kms\ (see above). 

It is tempting to interpret the high pressure as a result of 
ram pressure action of the halo plasma on the LMC gas. 
Using this concept, one can derive a density for the halo gas 
based on the X-ray luminosity (see Blondiau \ea 1997). 
Taking as velocity 465 \kms\ (up by a factor 2 from the one used by 
Blondiau et al.) leads to a halo density $n_{\rm e} \sim 6\ 10^{-3}$ \cmc. 
With temperatures between 10$^6$ and 10$^7$ K in the galactic halo the 
relative speed between LMC and halo is supersonic 
(sound speed is $\sim$100 \kms) so that a shock may be present indeed. 
The shock induced by the ram pressure causes the locally high pressure. 
Most importantly, the high pressure X-ray emitting area is 
located near the leading edge of the LMC 
and is well aligned with and directly adjacent to the 
large molecular cloud complex (Cohen \ea 1988) south of 30\,Dor. 
It appears very plausible that these clouds are 
- or will be in the near future - forming stars at a very high rate.

\begin{figure}
\def\epsfsize#1#2{1.0\hsize}
\centerline{\epsffile{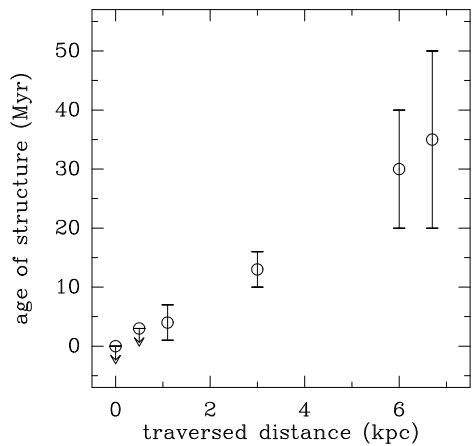}}
\caption[]{Ages of superstructures along the eastern and northern 
edge of the LMC are plotted against the distance they have rotated 
away from the location of the LMC bow-shock (data from Table 1). 
The zero point of the distance is chosen in the brightest IRAS spot 
of the dark cloud. 
The correlation between travel time and age is evident, 
suggesting that star formation is triggered indeed 
at the leading edge of the LMC
} 
\end{figure}

\section{Evolution after the formation trigger}

When large gas complexes are compressed star formation is triggered. 
The stellar contraction process requires less 
than 10$^5$ yr for massive stars 
and up to 5 Myr for a star of $\simeq$3 \ms\ (Bernasconi \& Maeder 1996). 
Stars at the top of the main sequence will start to evolve 
even before the stars of low mass have reached the main sequence. 

The brightest superstructure will be the one where the massive 
O and B stars still produce their gigantic amounts of Lyman continuum photons. 
The gas still present locally between the stars is thus 
very luminous in H$\alpha$. 
This happens between 5 and 10 Myr after the star forming burst. 
The delay is due to the time needed to get the birth cloud ionized 
in the first place. 
The calculation is along the same lines as in Braun \ea (1997) for 
the determination of the time dependent evolution of the supernova rate. 

After some 5 Myr the first stars will turn supernova. 
The supernova rate will increase considerably over the next few Myr 
(Braun \ea 1997) 
and these supernovae may blow the birth cloud apart, 
exposing the original association. 
This will have happened some 10-15 Myr after the first stars came into being.

\section{Travel distance fits age of star forming regions}

After the LMC bow-shock has triggered the star formation 
at the location of the largest impact, 
these star forming regions will move away from the 
leading edge due to the rotation of the LMC. 
In that case we should be able to relate the ages of the star groups 
with the LMC rotation velocity of $\sim$ 150 \kms. 

Starting at the leading edge in the SE we find clockwise:
the giant dark cloud strong in far-infrared emission; the very young 
star cluster in N\,159 at the northern part of the dark cloud; 
the 30\,Dor complex; 
the supergiant shell LMC~4. 
The ages of these structures (Table 1) are clearly sequential in time. 
Also the age of NGC 1818 and of the surrounding field fit in this sequence. 

We then measured the distances between these superstructures 
as seen projected on the sky and then accounted for the LMC inclination. 

The data are plotted in Fig.\,1. 
The slope of the line connecting the entries in Fig.\,1 gives the 
velocity of rotation. 
Given the uncertainty in the data points, which is mostly in the age 
(a parameter normally having errors symmetric in log age),
we find a rotation of 130 - 200 \kms. 
This value is in line with our considerations of Sect.\,4, 
demonstrating our proposition.

\section{Predictions from the model}

The dark cloud in the SE most likely contains protostars, 
which may be observable in the infrared. 
Photometry with current IR-sensitive CCD arrays should be able to uncover 
many embedded stars. 

Since the star formation was triggered at the edge of the LMC, 
and since the rotation will move these structures away but will keep 
them at the outer edge of the LMC, 
the energy put into the environment will start to escape in all directions, 
but least toward the inner side of the LMC disk. 
We thus will see these structures open up toward the outer edge. 
This is clearly the case for LMC~4, 
but also for the new shell, as proposed by Kim \ea (1997) 
based on radiosynthesis observations with the Australia Telescope. 

We may expect in the NW of the LMC a superstructure 
with an age of about 40 Myr. 
N\,11, the very bright \hii\ region has an age of about 10 Myr 
(Walborn \& Parker 1992).
Here we note the finding by Cole \ea (1997) of a 
relatively young field star population of about 20-50 Myr 
near the globular cluster NGC 1783 ($\sim$20\arcmin\ to the NNE of N\,11). 
The SERC-J plate shows in that area a large association of blue stars. 
N\,11 contains the secondary generation, 
much like NGC 1948 does at the edge of LMC\,4 (Vallenari \ea 1993). 
The age and distance for the field around NGC\,1783 are included in 
Table 1 and Fig.\,1.

A possible superstructure in the SW corner of the LMC should, if the 
bow-shock starformation relation with rotation is correct, 
have an age of 70-100 Myr.

\section{Concluding remarks}

The visual dominance of 30 Dor tempts us 
to think that this structure is the `centre' of the LMC. 
This notion is not correct, as argued above. 
It is well known that most of the mass of a galaxy 
is stored in the low mass stars, 
whereas most of the light (star light and H$\alpha$) 
comes from the high mass stars and their vicinity. 
As all starforming galaxies do, the LMC puts on a show, 
tempting us away from its real structure. 
The `activity centre' of the LMC rather lies in the SE, 
where the bow-shock is essential in triggering star formation. 
However, we do not exclude that sequential star formation 
after a big triggering event produces further young stars, 
nor that other star formation mechanisms 
have been or are active elsewhere in the LMC.

\acknowledgements
This research is carried out in the framework of the Bochum/Bonn 
Graduiertenkolleg `The Magellanic Clouds and other Dwarf Galaxies' 
of the Deutsche Forschungsgemeinschaft (DFG). 
A.V. thanks the Graduiertenkolleg for supporting a research stay in Bonn. 
We thank present and former members of the Graduiertenkolleg 
for frequent discussions.
We thank the referees to point out a description of star formation 
by the LMC bow-shock by Dopita (1987).

\end{document}